\documentclass[11pt]{article}
\usepackage{amsfonts,amsmath}

\begin{document}

\title{\textbf{On boundary conditions }\\
\textbf{in three-dimensional AdS gravity}}
\author{Olivera Mi\v{s}kovi\'{c} $^{a,b}$ and Rodrigo Olea $^{b,c}
\medskip $ \\
$^a${\small \emph{Instituto de F\'{\i}sica, P. Universidad
Cat\'{o}lica de Valpara\'{\i}so, Casilla 4059, Valpara\'{\i}so, Chile.}}\\
$^b${\small \emph{Departamento de F\'{\i}sica, P. Universidad
Cat\'{o}lica de Chile, Casilla 306, Santiago 22,
Chile.}}\\
$^c${\small \emph{Centro Multidisciplinar de Astrof\'{\i}sica -
CENTRA,\ Departamento de F\'{\i}sica,}}\\
{\small \emph{\ Instituto Superior T\'{e}cnico, Universidade T\'{e}cnica de
Lisboa, }} \\
{\small \emph{Av. Rovisco Pais 1, 1049-001 Lisboa, Portugal.}}\\
 {\small E-mail: \texttt{{olivera.miskovic@ucv.cl\thinspace ,
rolea@\texttt{fisica.ist.utl.pt}}}}}
%\vspace*{5mm}
%PACS numbers:\quad 04.20.Cv, 04.20.Ha,\ 04.60.Kz\qquad}
\date{}
\maketitle

\begin{abstract}
A finite action principle for three-dimensional gravity with
negative cosmological constant, based on a boundary condition for
the asymptotic extrinsic curvature, is considered. The bulk action
appears naturally supplemented by a boundary term that is one half
the Gibbons-Hawking term, that makes the Euclidean action and the
Noether charges finite without additional Dirichlet counterterms.
The consistency of this boundary condition with the Dirichlet
problem in AdS gravity and the Chern-Simons formulation in three
dimensions, and its suitability for the higher odd-dimensional
case, are also discussed.
\end{abstract}

\section{Introduction}

Three-dimensional gravity with negative cosmological constant
\cite{Deser-Jackiw} is a simple model that catches the main
features present in $D>3$ dimensions. In fact, this theory --first
considered by Deser and Jackiw in \cite{Deser-Jackiw}-- has black
hole solutions, possesses a rich asymptotic dynamics and, as in
the higher-dimensional case, its action also needs to be
regularized in order to give rise to finite conserved charges and
Euclidean action.

The dynamics at the boundary is determined by the asymptotic behavior of the
gravity fields. Supplementing the action with appropriate boundary terms and
demanding boundary conditions, the asymptotic dynamics of three-dimensional
AdS gravity is described by a Liouville theory \cite{CvDH}.

The boundary dynamics is essential for a well-posed definition of
the global charges. For example, the algebra of asymptotically
locally AdS gravity in three dimensions is infinite-dimensional
conformal algebra described by Virasoro generators, whose
Hamiltonian realization in terms of conserved charges introduces a
non-trivial classical central charge \cite{Brown-Henneaux}.

Many interesting properties of this gravity theory are due to the
fact that the AdS gravity can be formulated as Chern-Simons theory
for $SO(2,2)$ group \cite{Achucarro-Townsend} (see also
\cite{Witten88}). In this context, the global charges in
Chern-Simons AdS gravity in Hamiltonian formalism were studied in
\cite{Banados}.

In the framework of AdS/CFT correspondence
\cite{Maldacena,Witten}, the duality between AdS gravity and a
Conformal Field Theory on the boundary is realized by the
identification between the gravitational quasilocal stress tensor
and the conformal energy-momentum tensor. In that way, the stress
tensor in the CFT generating functional couples to the boundary
metric (initial data for the Einstein equation), from where the
$n$-point functions are computed. In the AdS gravity side, this
information is encoded in the finite part of the stress tensor,
that needs to be regularized by a procedure that respects general
covariance on the boundary (holographic renormalization)
\cite{Sk-He}. This method provides an algorithm to construct the
(Dirichlet) counterterms to achieve finite conserved quantities
and Euclidean action (see, e.g.,
\cite{Ba-Kr,Emparan-Johnson-Myers,Pa-Sk2,Mann}).

 In practice, however, this regularization procedure
is easy to carry out only for low enough dimensions, because the
number of possible counterterms increase drastically with the
dimension. Moreover, these terms do not seem to obey any
particular pattern and the full series for an arbitrary dimension
is still unknown.

  An alternative to this construction of boundary terms
was proposed in \cite{MOTZodd} for odd dimensions and
\cite{OleaJHEP} for even dimensions, where the boundary terms have
a geometrical origin (closely related to Chern-Simons forms), and
that is based on boundary conditions that are not the standard
Dirichlet one. For instance, even in $D=4$, a different boundary
condition leads to a boundary term that regularizes the AdS
action, but that does not recover the Gibbons-Hawking term plus
Dirichlet counterterms, as a consequence of a different finite
action principle. But, at the same time, this boundary term is
dictated by the Euler theorem, showing the profound connection
with topological invariants.

This paper understands the simplest example of the odd-dimensional
regularization scheme proposed in \cite{MOTZodd}. Even though the
explicit relation to the Dirichlet problem is possible here, a
comparison in the general case is still unknown.
 The guideline to achieve finite
conserved charges and Euclidean action is a well-defined action
principle for a boundary condition on the extrinsic curvature. In
spite of the simplicity of 3D, the suitability of this boundary
condition for higher odd-dimensional gravity becomes evident from
its compatibility with the Dirichlet problem in AdS gravity.

\section{The action principle}

We consider three-dimensional AdS gravity described by the action%
\begin{equation}
I=-\frac{1}{16\pi G_{N}}\left[ \int_{M}d^{3}x\,\sqrt{-G}\left( \hat{R}+\frac{%
2}{\ell ^{2}}\right) +2\alpha \int_{\partial M}d^{2}x\,\sqrt{-h}K\right] \,,
\label{Ig3tensor}
\end{equation}%
where $\ell $ is the AdS radius and we have supplemented the bulk
Lagrangian by a boundary term that is $\alpha $ times the
Gibbons-Hawking term \cite{Gibbons-Hawking}.

As it is standard in holographic renormalization \cite{Sk-He}, we
take a Gaussian (normal) form for the spacetime metric%
\begin{equation}
ds^{2}=G_{\mu \nu}dx^{\mu} dx^{\nu} = N^{2}(\rho )\,d\rho
^{2}+h_{ij}(\rho ,x)dx^{i} dx^{j} \,, \label{radialADM}
\end{equation}%
such that the only relevant boundary is at $\rho =const.$
However, we shall not take any particular expansion for the boundary metric $%
h_{ij}(\rho ,x)$.

We will work in the language of differential forms, with the
dreibein $e^{A}=e_{\mu }^{A}\,dx^{\mu }$ (the spacetime metric is
$G_{\mu \nu }=\eta _{AB}\,e_{\mu }^{A}e_{\nu }^{B}$) and the spin
connection $\omega ^{AB}=\omega _{\mu }^{AB}\,dx^{\mu }$ because
certain features of the theory become manifest in terms of
differential forms, as we shall see below.

In order to preserve the Lorentz covariance of the boundary term,
we introduce the second fundamental form (SFF) as the difference
between the dynamical field $\omega ^{AB}$ and a fixed spin
connection $\bar{\omega}^{AB}$,
\begin{equation}
\theta ^{AB}=\omega ^{AB}-\bar{\omega}^{AB}\,.  \label{SFFdefin}
\end{equation}
For the gauge (\ref{radialADM}), the dreibein adopts the block form $e^{1}=Nd\rho $ and $%
e^{a}=e_{i}^{a}\,dx^{i}$ with the indices splitting $A=\{1,a\}$.
The spin connection decomposes as $\omega ^{AB}=\{\omega
^{1a},\omega ^{ab}\}$. For the torsionless case, the block $\omega
^{ab}$ is related to the Christoffel symbol
$\hat{\Gamma}_{ij}^{k}(G)=\Gamma _{ij}^{k}(h)$ of the boundary
metric $h_{ij}$, that transforms as a connection (and not a
tensor) in the boundary indices, so that it cannot enter the
boundary term explicitly. On the other hand, for the rest of the
components on $\partial M$, we have
\begin{equation}
\omega ^{1a}=K_{i}^{j}\,e_{j}^{a}\,dx^{i}=K^{a}\,,  \label{normalomega}
\end{equation}%
where the extrinsic curvature $K_{ij}$ in normal coordinates (\ref%
{radialADM}) is given by%
\begin{equation}
K_{ij}=N\,\hat{\Gamma}_{ij}^{\rho }=-\frac{1}{2N}\,\partial _{\rho }h_{ij}.
\label{Kdef}
\end{equation}%
The explicit dependence on $\omega ^{ab}$ can be removed by taking
$\bar{\omega}^{AB}$ as coming from a product metric
\begin{equation}
ds^{2}=\bar{N}^{2}(\rho )\,d\rho ^{2}+\bar{h}_{ij}(x)\,dx^{i}dx^{j}
\label{coborh}
\end{equation}
cobordant to the dynamical one, i.e., it matches $h_{ij}$ only on
the boundary, $\bar{h}_{ij}(x)=h_{ij}(\rho _{0},x)$ and such that
this spin connection on $\partial M$ contains only tangential
components \cite{eguchi,Spivak},
\begin{equation}
\bar{\omega}^{1a}=0\,,\qquad \bar{\omega}^{ab}=\omega ^{ab}.
\label{omegabar}
\end{equation}
Thus, the SFF can be used to express all the quantities as
boundary tensors (e.g., the extrinsic curvature),
\begin{equation}
\theta ^{1a}=K_{i}^{a}\,dx^{i},\qquad \theta ^{ab}=0\,.  \label{thetanormal}
\end{equation}

The explicit form taken by the SFF in normal coordinates
(\ref{radialADM}) is the key point to obtain the boundary term in
the Euler theorem in four dimensions \cite{eguchi}. This argument
has also been used to obtain the boundary term that regularizes
AdS gravity in higher odd \cite{MOTZodd} and even \cite{OleaJHEP}
dimensions.

With the above definitions, the action (\ref{Ig3tensor}) can be
written
\begin{equation}
I=\frac{1}{16\pi G_{N}}\left[ \int_{M}\varepsilon _{ABC}\left( \hat{R}^{AB}+%
\frac{1}{3\ell ^{2}}\,e^{A}e^{B}\right) e^{C}-\alpha
\int_{\partial M}\varepsilon _{ABC}\,\theta ^{AB}e^{C}\right] \,,
\label{Ig3forms}
\end{equation}
in terms of the Lorentz curvature
$\hat{R}^{AB}=\frac{1}{2}\,\hat{R}_{\mu \nu }^{AB}\,dx^{\mu
}\wedge dx^{\nu }=d\omega ^{AB}+\omega _{C}^{A}\wedge \omega
^{CB}$, the SFF, the triad and the Levi-Civita tensor, defined as
$\varepsilon _{012}=-1$. We omit the wedge product between
differential forms.

An arbitrary variation of this action, projected in the frame
(\ref{radialADM}), produces the surface term
\begin{equation}
\delta I=-\frac{1}{8\pi G_{N}}\int_{\partial M}\varepsilon
_{ab}\left[ (1-\alpha )\delta K^{a}e^{b}-\alpha K^{a}\delta
e^{b}\right] \,, \label{deltaI3}
\end{equation}%
when equations of motion hold. The Levi-Civita tensor in two
dimensions is defined as $\varepsilon _{ab}=-\varepsilon _{1ab}$.
We also used the fact that any variation
acting on the SFF is $\delta \theta ^{AB}=\delta \omega ^{AB}$, as $\bar{%
\omega}^{AB}$ is kept fixed on the boundary $\partial M$.

In a radial foliation of the spacetime (\ref{radialADM}) the
boundary metric and the extrinsic curvature are independent
variables. In fact, $K_{ij}$ is closely related to the conjugate
momentum of $h_{ij}$, where the radial coordinate plays the role
of time. Standard choice $\alpha =1$ clearly recovers the
Gibbons-Hawking term and defines the Dirichlet problem for
gravity, because it eliminates the
variation of $K^{a}$ and replaces it by a variation of the boundary dreiben $%
e^{b}$, producing the surface term%
\begin{equation}
\delta I_{D}=\frac{1}{16\pi G_{N}}\int_{\partial M}d^{2}x\,\sqrt{-h}\left(
K^{ij}-h^{ij}K\right) \delta h_{ij}\,.  \label{alpha1-2}
\end{equation}%
This choice of $\alpha $ ensures a well-posed action principle for arbitrary
variations of the boundary metric $h_{ij}$. However, the action $I_{D}$
requires a counterterm%
\begin{equation}
I_{reg}=I_{D}+\frac{1}{8\pi G_{N}}\int_{\partial M}d^{2}x\,\frac{1}{\ell }\,%
\sqrt{-h}\,,
\end{equation}%
to achieve the finiteness of both the Euclidean action and the conserved
quantities \cite{Ba-Kr} obtained through a quasilocal (boundary) stress
tensor definition \cite{Brown-York}.

Here, we shall consider a different coefficient $\alpha =1/2$ and
analyze the consequences of this choice. As it can be seen from
Eq.(\ref{deltaI3}), the surface term takes the form
\begin{equation}
\delta I=-\frac{1}{16\pi G_{N}}\int_{\partial M}\varepsilon
_{ab}\left( \delta K^{a}e^{b}-K^{a}\delta e^{b}\right)
\label{curl3}
\end{equation}
that, with the help of $\delta K^{a}=\left( \delta
K_{i}^{j}e_{j}^{a}+K_{i}^{j}\delta e_{j}^{a}\right) dx^{i}$, can
be written as
\begin{equation}
\delta I=-\frac{1}{16\pi G_{N}}\int_{\partial
M}d^{2}x\,\varepsilon _{ab}\varepsilon ^{ik}\left[ \delta
K_{i}^{j}e_{j}^{a}e_{k}^{b}+\delta e_{j}^{a}e_{l}^{b}\left(
K_{i}^{j}\delta _{k}^{l}-K_{k}^{l}\delta _{i}^{j}\right) \right] .
\label{curl3tensor}
\end{equation}%
In this case, the action becomes stationary only under a suitable boundary
condition on the extrinsic curvature $K_{i}^{j}$.

\section{Asymptotic conditions}

We consider fixing the extrinsic curvature on the boundary
$\partial M$, that is,
\begin{equation}
\delta K_{i}^{j}=0\,,\label{deltaK0}
\end{equation}%
in order to cancel the first term in Eq.(\ref{curl3tensor}). This means
that, in the asymptotic region, $K_{i}^{j}$ tends to a (1,1)-tensor with
vanishing variation. For simplicity, we take%
\begin{equation}
K_{i}^{j}=\frac{1}{\ell }\,\delta _{i}^{j}\,,  \label{Kdelta}
\end{equation}%
where the $1/\ell $ factor is introduced in order to fix the scale for
asymptotically AdS (AAdS) spacetimes. This choice makes the rest of the
surface term in Eq.(\ref{curl3tensor}) vanish identically, so that the
gravitational action has indeed an extremum for that boundary condition.

To further understand the meaning of the condition (\ref{Kdelta}), we can
put Eq.(\ref{Kdef}) in the form%
\begin{equation}
K_{ij}=-\frac{1}{2}\,n^{\mu }\partial _{\mu }h_{ij}=-\frac{1}{2}\,\mathcal{L}%
_{n}h_{ij}\,,  \label{KLie}
\end{equation}%
where $\mathcal{L}_{n}$ is a directional (Lie) derivative along a unit
vector normal to the boundary, $n_{\mu }=(0,N,0)$. Inserting the definition (%
\ref{KLie}) in the asymptotic condition (\ref{Kdelta}), we see that the
latter relation is satisfied in a spacetime whose boundary $\partial M$ is
endowed with a conformal Killing vector because%
\begin{equation}
\mathcal{L}_{n}h_{ij}=\hat{\nabla}_{i}n_{j}+\hat{\nabla}_{j}n_{i}=-\frac{2}{%
\ell }\,h_{ij}\,.
\end{equation}%
A submanifold whose extrinsic curvature is proportional to the induced
metric is usually referred to as \emph{totally umbilical} \cite{Spivak}.

In order to describe AAdS spacetimes, it is common to take the
lapse function as $N=\ell /2\rho $ and the boundary metric as
$h_{ij}(\rho ,x)=g_{ij}(\rho ,x)/\rho $, so that
\begin{equation}
ds^{2}=\frac{\ell ^{2}}{4\rho ^{2}}\,d\rho ^{2}+\frac{1}{\rho }%
\,g_{ij}\left( \rho ,x\right) dx^{i}dx^{j}\,,  \label{confcoord}
\end{equation}%
that is suitable to represent the conformal structure of the
boundary located at $\rho =0$. According to Fefferman and Graham
\cite{F-G}, the metric $g_{ij}\left( \rho ,x\right) $ is regular
on the boundary and it can be expanded around $\rho=0$ as
\begin{equation}
g_{ij}\left( \rho ,x\right) =g_{(0)ij}\left( x\right) +\rho
g_{(1)ij}\left( x\right) +\rho^{2}g_{(2)ij}\left( x\right) +\cdots
\,, \label{FGexp}
\end{equation}
where $g_{(0)ij}$ is a given initial data for the metric.  In
three dimensions, the Weyl tensor vanishes identically and the FG
series (\ref{FGexp}) becomes finite, terminating at order $\rho^2$
\cite{Skenderis-Solodukhin}. The solution of the Einstein equation
in this case is $g_{(2)ij}={1 \over 4}\,
(g_{(1)}g_{(0)}^{-1}g_{(1)})_{ij}$, where $g_{(1)ij}$ has the
trace fixed in terms of the curvature of $g_{(0)ij}$.

The standard Dirichlet boundary condition on $h_{ij}$ is in
general ill-defined for AdS gravity because of its conformal
boundary. Indeed, it follows from its asymptotic form
(\ref{confcoord},\ref{FGexp}) that the induced metric is divergent
at the boundary and therefore,  it is not suitable to fix it
there. Alternatively, one can demand that a conformal structure
(i.e., its representative $g_{(0)ij}$) is kept fixed at the
boundary. As discussed in \cite{Pa-Sk2}, this action principle
requires the addition of new boundary terms apart from the usual
Gibbons-Hawking term. However, it can be proven that these extra
terms are indeed the usual Dirichlet counterterm series.

The compatibility of the boundary condition (\ref{Kdelta}) with
the Fefferman-Graham form of the metric
(\ref{confcoord},\ref{FGexp}) is then
evident from the expansion of the extrinsic curvature (\ref{Kdef}),%
\begin{equation}
K_{i}^{j}=\frac{1}{\ell }\,\delta _{i}^{j}- \frac{\rho}{\ell }%
\,g_{(1)ik}\,g_{(0)}^{kj}+\cdots \,,  \label{KFG}
\end{equation}%
that contains only increasing powers of $\rho $. This also implies
that fixing the extrinsic curvature at the boundary is equivalent
to keeping fixed the conformal structure. As a consequence, the
Dirichlet problem for the conformal metric as the boundary data
can be converted into the initial-value problem for $K_{ij}$, such
that the standard holographic renormalization can be reformulated
in terms of the extrinsic curvature \cite{Pa-Sk}.

 As we shall see below, in the present case the regularization is encoded in the
 boundary term that extremizes the action for the boundary
 condition (\ref{Kdelta}).

\section{Regularized action}

For $\alpha =1/2$, the action is written as%
\begin{equation}
I=-\frac{1}{16\pi G_{N}}\left[ \int_{M}d^{3}x\,\sqrt{-G}\left( \hat{R}+\frac{%
2}{\ell ^{2}}\right) +\int_{\partial M}d^{2}x\,\sqrt{-h}K\right] .
\label{Ig3onehalf}
\end{equation}%
Its variation on-shell is the surface term (\ref{curl3}), containing both
variations of the boundary dreibein $e^{a}$ and the extrinsic curvature $%
K^{a}$. The formulation in terms of these variables is useful to
recover the conserved quantities displayed below from a generic
Chern-Simons theory in three dimensions.

The Noether current can be written as \cite{Ramond,Iyer-Wald}
\begin{equation} \label{Noether}
\ast J=-\Theta (e^{a},K^{a},\delta e^{a},\delta K^{a})-i_{\xi
}\left( L+dB\right) \,,
\end{equation}
where $\Theta $\ is the surface term in the variation of the
action (\ref{curl3}), $L$ and $B$ are the bulk Lagrangian and the
boundary term in Eq.(\ref{Ig3onehalf}), respectively, and $i_{\xi
}$\ is the contraction operator with the Killing vector $\xi ^{\mu
}$ \cite{ichi}. The Noether charges, with the contributions coming
from the bulk and the boundary, are then given by
\begin{eqnarray}
Q(\xi ) &=&\mathcal{K}(\xi )+\int_{\partial \Sigma }\left( i_{\xi
}K^{a}\,\frac{\delta B}{\delta K^{a}}+i_{\xi }e^{a}\,\frac{\delta
B}{\delta
e^{a}}\right)  \notag \\
&=&\mathcal{K}(\xi )-\frac{1}{16\pi G_{N}}\int_{\partial \Sigma }\varepsilon
_{ab}\left( i_{\xi }K^{a}e^{b}-K^{a}i_{\xi }e^{b}\right) \,.
\label{chargegeneral}
\end{eqnarray}
The first term is known as the Komar's integral%
\begin{equation}
\mathcal{K}(\xi )=\frac{1}{8\pi G_{N}}\int_{\partial \Sigma }\varepsilon
_{ab}\,i_{\xi }K^{a}e^{b}\,,
\end{equation}%
and it is the conserved quantity associated to the bulk term in
the gravity action.

Finally, the conserved quantities for three-dimensional AdS
gravity read
\begin{eqnarray}
Q(\xi ) &=& \frac{1}{16\pi G_{N}}\int_{\partial \Sigma
}\varepsilon _{ab}\left( i_{\xi }K^{a}e^{b}+K^{a}i_{\xi
}e^{b}\right) \notag\\
 &=& \frac{1}{16\pi
G_{N}}\int_{\partial \Sigma }\sqrt{-h}\,\varepsilon _{ij}\,\xi
^{k}\left( \delta _{l}^{j}K_{k}^{i}+\delta
_{k}^{j}K_{l}^{i}\right) dx^{l}. \label{Q3tens}
\end{eqnarray}

Stationary, circularly symmetric black holes exist in
three-dimensional gravity only in presence of negative
cosmological constant. The metric for the BTZ black hole
\cite{BTZ} reads
\begin{equation}
ds^{2}=-\gamma
(r)f^{2}(r)\,dt^{2}+\frac{dr^{2}}{f^{2}(r)}+r^{2}\left( d\varphi
+n(r)dt\right) ^{2},  \label{metricBH3}
\end{equation}
with
\begin{equation}
\!\!\!\!\!f^{2}(r) =-8G_{N}\,M+\frac{r^{2}}{\ell
^{2}}+\frac{16G_{N}^{2}J^{2}}{r^{2}}\,,\qquad
n(r) = -\frac{4G_{N}J}{r^{2}}\,\qquad\gamma(r)=1.  \label{defs} \\
\end{equation}
The horizon $r_{+}$ is defined by the largest radius satisfying
$f(r_{+})=0$.

For the isometries\ $\partial/ \partial _{t}$ and\ $\partial
/\partial _{\varphi }\,$, the charge formula (\ref{Q3tens})
provides the correct
conserved quantities for the BTZ metric,%
\begin{equation}
Q(\partial _{t}) = M\,,\qquad Q(\partial _{\varphi }) = J\,,  \label{mass3} \\
\end{equation}%
where $\partial \Sigma $ is taken as $S^{1}$ at radial infinity.
The vacuum energy for three-dimensional AdS space corresponds to $%
M=-1/8G_{N} $. On the contrary to the Hamiltonian approach
\cite{BTZ} or perturbative Lagrangian methods \cite{A-D}, we do
not need to specify the background to obtain the correct results
(\ref{mass3}).

%%%%%%%%%%%%%%%%%%%%%%%%%%%%%%%%%%%%%%%%%%%%%%%%%%%%%%%%%%%%%%%%%%%%%%
The \emph{regularized} action (\ref%
{Ig3onehalf}) does not lend itself for a clear definition of a
boundary stress tensor $T^{ij}$ because its variation
(\ref{curl3}) contains a piece along $\delta K^{a}$ that it is
usually cancelled by the Gibbons-Hawking term. However, we
can rewrite the action as%
\begin{equation}
I=I_{D}+\frac{1}{16\pi G_{N}}\int_{\partial
M}d^{2}x\,\sqrt{-h}\,K\,, \label{Imia}
\end{equation}%
where $I_{D}$ stands for the action suitable for the Dirichlet problem (Eq.(%
\ref{Ig3tensor}) with $\alpha =1$). We will consider now the extra term in (%
\ref{Imia}) as a functional of the boundary metric $h_{ij}(\rho
,x)=g_{ij}(\rho ,x)/\rho $. The extrinsic curvature (\ref{Kdef})
can be generically written as%
\begin{equation}
K_{i}^{j}=\frac{1}{\ell }\,\delta _{i}^{j}-\frac{\rho }{\ell
}\,k_{i}^{j}\,, \label{Kexpgen}
\end{equation}%
with $k_{i}^{j}=g^{jk}\partial _{\rho }g_{ki}$, so that the second
term in Eq.(\ref{Imia}) takes the form

\begin{equation}
\frac{1}{16\pi G_{N}}\,\sqrt{-h}K=\frac{1}{8\pi G_{N}\ell }\left( \sqrt{-h}-2%
\sqrt{-g}\,k\right) \,.  \label{newbt}
\end{equation}%
The first term is just the Balasubramanian-Kraus counterterm
\cite{Ba-Kr}, whereas the second one can be shown to be a
topological invariant of the boundary metric $g_{(0)}$, that is,
$\sqrt{-g_{(0)}}\,R_{(0)}$. This follows from the fact that
$-2\sqrt{-g}\,k=-2\sqrt{-g_{(0)}}\,$Tr$(g_{(1)})$ on the boundary.
Indeed, the 3D Einstein equation in the gauge (\ref{confcoord})
determines the trace and vanishing covariant divergence of
$g_{(1)ij}$ \cite{Sk-He}. Then, the boundary term in
Eq.(\ref{Ig3onehalf}) both regularizes
the quasilocal stress tensor and reproduces the correct Weyl anomaly \cite%
{Rooman}.

The above argument also explains why the Euclidean action
supplemented by a Gibbons-Hawking term with an \emph{anomalous}
factor is finite, as first noticed in \cite{Banados-Mendez} where
it correctly describes the thermodynamics of the BTZ black hole.
It has been shown in \cite{Pa-Sk2} that the counterterms
constructed in the regularization procedure in terms of the
extrinsic curvature \cite{Pa-Sk} (and that are equivalent to
standard counterterms) allows to prove the first law of black hole
thermodynamics for a general asymptotically AdS black hole. Here,
it follows from the equivalence of the boundary term in
(\ref{Ig3onehalf}) to the Dirichlet counterterms plus a
topological invariant that the right thermodynamics is recovered
in a general case.

\section{Chern-Simons formulation}

The boundary term in (\ref{Ig3onehalf}) arises naturally in the
Chern-Simons formulation of three-dimensional
AdS gravity \cite{Banados-Mendez}. Indeed, the Chern-Simons (CS) action%
\begin{equation}
I_{CS}\left[ A\right] =\frac{k}{4\pi }\int_{M}\text{Tr}\left( AdA+\frac{2}{3}%
\,A^{3}\right)  \label{ICS}
\end{equation}%
for the AdS group $SO(2,2)$ whose gauge connection is given by%
\begin{equation}
A_{AdS}=\frac{1}{2}\,\omega ^{AB}J_{AB}+\frac{1}{\ell }\,e^{A}P_{A}\,,
\label{AAdS}
\end{equation}
and the trace of the AdS generators set
Tr$(J_{AB}P_{A})=\varepsilon _{ABC}$, is equivalent to the
Einstein-Hilbert-AdS bulk action plus the Ba\~{n}ados-Mendez
boundary term
\begin{equation}
\frac{1}{16\pi G_{N}}\int_{\partial M}\omega
_{A}\,e^{A}=\frac{1}{16\pi G_{N}}\int_{\partial
M}d^{2}x\,\sqrt{-h}\,K  \label{B-M}
\end{equation}
in the coordinate frame (\ref{radialADM}) (here $\omega
_{A}=\frac{1}{2} \,\varepsilon _{ABC}\,\omega ^{BC}$). Clearly,
the boundary term (\ref{B-M}) is not Lorentz-covariant as the one
constructed up with the SFF in Eq.(\ref{Ig3forms}). This is an
accident that happens only in $(2+1)$ dimensions: the dreibein
cannot go along $d\rho $ at the boundary and so the boundary term
does not depend on $\omega ^{ab}$. Therefore, the residual $2D$
Lorentz symmetry on $\partial M$ permits to express (\ref{B-M}) as
tensors on the boundary (the metric $h_{ij}$ and the extrinsic
curvature $K_{ij}$). In an arbitrary local Lorentz frame, the
non-invariance of the boundary term under local Lorentz
transformations produces extra asymptotic degrees of freedom
responsible for the arbitrary
coupling constant $\lambda $ in the potential term of Liouville theory \cite%
{Carlip}, that is either zero or put by hand in a metric
formulation.

In higher odd dimensions, one can also pass from the CS
formulation for the AdS group $SO(2n,2)$ in terms of the
connection $A$ to a Lovelock-type Lagrangian for gravity, i.e., a
polynomial in the Riemmann two-form and the metric
\cite{chamseddine}. In doing so, however, the produced boundary
term will be neither Lorentz-covariant nor the correct one that
regulates the conserved quantities and the Euclidean action for CS
black holes \cite{DCBH}. The introduction of the SFF is then
essential to restore Lorentz covariance and it also provides a
clear guideline for its explicit construction \cite{MOTZCS}.

Usually, the Chern-Simons formulation for $SO(2,2)$ exploits the fact that
the AdS gravity action can be written as the difference of two copies of the
CS action (\ref{ICS}) for $SO(2,1)$ \cite{Witten88} (in the Euclidean case, $%
SL(2,\mathbb{C})$)%
\begin{equation}
I=I_{CS}\left[ A\right] -I_{CS}\left[ \bar{A}\right] \,,  \label{I2copies}
\end{equation}%
where the connections for each copy of $SO(2,1)$ are%
\begin{equation}
A^{A}=\omega ^{A}+\frac{1}{\ell }\,e^{A},\qquad \bar{A}^{A}=\omega ^{A}-%
\frac{1}{\ell }\,e^{A},  \label{AAAdS}
\end{equation}%
with $k=-\ell /4G_{N}$.

The variation of the action (\ref{I2copies}) produces the
equations of motion plus a surface term that is cancelled by
taking \emph{chiral} boundary conditions
\begin{equation}
A_{\bar{z}}=0\quad \text{and\quad }\bar{A}_{z}=0\,,  \label{chiralbc}
\end{equation}
with the use of the light-cone coordinates $z=t+\ell \varphi $ and $\bar{z}%
=t-\ell \varphi $ for Minkowskian signature \cite{CvDH} (the
Euclidean version considers the same set of boundary conditions,
but for complex coordinates $(z,\bar{z})$ defined on the solid
torus that describes the topology of the Euclidean black hole
\cite{Banados-Mendez, MB3D}). In the Chern-Simons formulation, the
explicit form of the conditions (\ref{chiralbc}) is
\begin{eqnarray}
2\ell A_{\bar{z}}^{A} &=&\left( \ell \,\omega _{t}^{A}-\frac{1}{\ell }%
\,e_{\varphi }^{A}\right) -\left( \omega _{\varphi }^{A}-e_{t}^{A}\right)
=0\,,  \label{Abarz} \\
2\ell \bar{A}_{z}^{A} &=&\left( \ell \,\omega _{t}^{A}-\frac{1}{\ell }%
\,e_{\varphi }^{A}\right) +\left( \omega _{\varphi }^{A}-e_{t}^{A}\right)
=0\,,  \label{barAz}
\end{eqnarray}
that are satisfied by AdS gravity. Indeed, the three-dimensional
black hole has
\begin{equation}
\begin{array}{ccc}
e^{0}=fdt\,,\qquad & e^{1}=\dfrac{1}{f}\,dr\,,\qquad & e^{2}=rN^{\varphi
}dt+rd\varphi \,,\medskip \\
\omega ^{0}=fd\varphi \,,\qquad & \omega ^{1}=\dfrac{J}{2r^{2}f}\,dr\,,\qquad
& \omega ^{2}=rN^{\varphi }d\varphi +\dfrac{r}{\ell ^{2}}\,dt\,.%
\end{array}
\label{bcBTZ}
\end{equation}

Thus, the action has an extremum for the chiral boundary
conditions, eqs.(\ref{Abarz},\ref{barAz}), that is clearly not the
standard Dirichlet one for the metric.  This shows that there are
(at least) two ways of regularizing the AdS gravity in three
dimensions. In this paper, we consider another boundary condition
that also explains the anomalous factor in the Gibbons-hawking
term.

On the contrary to the relations (\ref{Abarz},\ref{barAz}) fulfilled by the
bulk geometry, the boundary condition (\ref{Kdelta}) implies%
\begin{equation}
\omega _{i}^{a1}=\frac{1}{\ell }\,e_{i}^{a}  \label{mybc}
\end{equation}%
only in the asymptotic region ($a=\{0,2\}$ and $i=\{t,\varphi \}$). In fact,
in the CS formulation of 3D AdS gravity, the surface term coming from an
arbitrary variation is%
\begin{eqnarray}
\delta I_{CS}\left[ A_{AdS}\right] &=&-\frac{k}{4\pi }\int_{\partial M}\text{%
Tr}\left( A\delta A\right)  \label{varICSA} \\
&=&\frac{1}{32\pi G_{N}}\int_{\partial M}\varepsilon _{ABC}\left( \delta
\omega ^{AB}e^{C}-\omega ^{AB}\delta e^{C}\right) \,,  \label{varICSomegae}
\end{eqnarray}%
that reduces to Eqs.(\ref{curl3},\ref{curl3tensor}) for the radial
foliation (\ref{radialADM}).

The AdS connection can also be written as $A=\frac{1}{2}\,W^{\bar{A}\bar{B}%
}J_{\bar{A}\bar{B}}$ using covering space indices $\bar{A}=\{A,3\}$, where $%
W^{AB}=\omega ^{AB}$ and $W^{A3}=\frac{1}{\ell }\,e^{A}$. Then, the
asymptotic condition (\ref{mybc}) adopts the compact form%
\begin{equation}
W_{i}^{a1}=W_{i}^{a3}.
\end{equation}
This new condition might have nontrivial consequences at the level
of the induced theory at the boundary.

The surface term (\ref{varICSomegae}) was obtained in \cite{clement} for the
Palatini form of the AdS gravity action%
\begin{equation}
\delta I_{g}=\frac{1}{32\pi G_{N}}\int_{\partial M}n_{\mu }\left[ \left(
\hat{\Gamma}_{\nu \lambda }^{\lambda }\delta \mathcal{G}^{\mu \nu }-\hat{%
\Gamma}_{\nu \lambda }^{\mu }\delta \mathcal{G}^{\nu \lambda }\right)
-\left( \mathcal{G}^{\mu \nu }\delta \hat{\Gamma}_{\nu \lambda }^{\lambda }-%
\mathcal{G}^{\nu \lambda }\delta \hat{\Gamma}_{\nu \lambda }^{\mu }\right) %
\right] \,,
\end{equation}%
where $\mathcal{G}^{\mu \nu }=\sqrt{-G}G^{\mu \nu }$. Clearly, the action
has an extremum for a mixed Dirichlet-Neumann boundary condition. However,
no clear identification of such boundary condition in terms of tensorial
quantities defined on $\partial M$ was made in this reference.

In CS formulation it is simple to obtain the Noether charges. With the
surface term (\ref{varICSA}) in terms of the gauge connection, we can
compute the conserved charges associated to an asymptotic Killing vector $%
\xi $ using the Noether theorem. The conserved current for this
theory is
\begin{equation}
\ast J=\frac{k}{4\pi }\left[ -\text{Tr}\left( A\mathcal{L}_{\xi }A\right)
-i_{\xi }\text{Tr}\left( AF-\frac{1}{3}A^{3}\right) \right] \,.  \label{JA}
\end{equation}%
The Lie derivative for a connection field takes the form {\normalsize $%
\mathcal{L}_{\xi }A=D(i_{\xi }A)+i_{\xi }F$}, where the covariant derivative
is defined as $D(i_{\xi }A)=d(i_{\xi }A)+[A,i_{\xi }A]$. Using the equation
of motion $F=dA+A^{2}=0$, and integrating by parts, the current is finally
expressed as%
\begin{equation}
\ast J=\frac{k}{4\pi }\,d\text{Tr}\left( AI_{\xi }A\right) \,,
\end{equation}%
from where we can read the conserved charge as \cite%
{Julia-Silva,chinos}%
\begin{equation}
Q(\xi )=\frac{k}{4\pi }\int\limits_{\partial \Sigma }\text{Tr}\left( Ai_{\xi
}A\right) \,.  \label{QCS3}
\end{equation}%
It is not difficult to prove that this expression recovers the
formula (\ref{Q3tens}) for the trace of AdS generators and the
radial foliation considered above.

\section{Conclusions}

In this paper, we show how a single boundary term solves at once
three problems in three-dimensional AdS gravity: it defines a
well-posed variation of the action (the action has extremum
on-shell under the condition (\ref{Kdelta})), produces finite
charges and regularizes the Euclidean action. In other words, the
Dirichlet counterterm is built-in in a boundary term that is 1/2
of the Gibbons-Hawking term.

A boundary condition on the extrinsic curvature (\ref{Kdelta}),
equivalent to keeping fixed the metric of the conformal boundary
in AAdS spacetimes, ensures a finite action principle in agreement
to the Dirichlet counterterms problem.

The same boundary condition leads to a regularization scheme of
AdS gravity alternative to the standard counterterms procedure
\cite{Sk-He,Ba-Kr,Emparan-Johnson-Myers,Sk,Odintsov}. In the new
approach, the boundary terms can depend, apart from intrinsic
quantities constructed out of the boundary metric $h_{ij}$ and
boundary curvature $R_{ij}^{kl}$, also on the extrinsic curvature
$K_{ij}$. At first, one might think that the number of possible
counterterms constructed up with these tensors is even higher than
in the standard procedure. However, the boundary condition
(\ref{Kdelta}) and another one on the asymptotic curvature --that
is identically satisfied in FG frame--  are restrictive enough to
substantially reduces the counterterms series to a compact
expression \cite{MOTZodd,OleaJHEP}.
 For
instance, in five dimensions the boundary term that regularizes
the AdS action is
\begin{equation}
I=-\frac{1}{16\pi G_N}\int_{M}d^{5}x\,\sqrt{-G}\,(\hat{R}-2\Lambda
)+c_{4}\int_{\partial M}B_{4} \,, \label{EHtensor}
\end{equation}
with $c_4=const.$ and the boundary term given by the expression
\begin{equation}
B_{4}=-\frac{1}{2}\,\sqrt{-h}\,\delta _{\left[ j_{1}j_{2}j_{3}j_{4}\right] }^{%
\left[ i_{1}i_{2}i_{3}i_{4}\right] }K_{i_{1}}^{j_{1}}\delta
_{i_{2}}^{j_{2}}\left(
R_{i_{3}i_{4}}^{j_{3}j_{4}}(h)-K_{i_{3}}^{j_{3}}K_{i_{4}}^{j_{4}}+\frac{1}{%
3\ell ^{2}}\,\delta _{i_{3}}^{j_{3}}\delta _{i_{4}}^{j_{4}}\right)
.
\end{equation}
It is worthwhile noticing that $B_{4}$ contains a term
proportional to $\sqrt{-h}\,K$, with a numerical factor that again
differs from the one of the Gibbons-Hawking term. The variation of
the above action --on-shell-- takes the form
\begin{eqnarray}
\delta I &=&2\int_{\partial M}\varepsilon _{abcd}\,\delta
K^{a}e^{b}\left[ \kappa\,
e^{c}e^{d}+c_{4}\left(\hat{R}^{cd}+\frac{1}{3\ell
^{2}}\,e^{c}e^{d}\right)\right]
\nonumber \\
&&-\frac{c_{4}}{2}\,\varepsilon _{abcd}\left( \delta
K^{a}e^{b}-K^{a}\delta e^{b}\right) \left(
R^{cd}-\frac{1}{2}K^{c}K^{d}+\frac{1}{2\ell
^{2}}\,e^{c}e^{d}\right)\,,
\end{eqnarray}%
where $\kappa =1/(96\pi G_N)$ and $\hat{R}^{cd}=R^{cd}-K^{c}K^{d}$
is the Gauss-Coddazzi relation for the Riemann tensor. An
appropriate choice of the coupling constant, $c_{4}=3\kappa \ell
^{2}/2$, makes the first line of above equation proportional to
the AdS curvature $\hat{R}^{cd}+\frac{1}{\ell ^{2}}\,e^{c}e^{d}$,
that vanishes at the boundary for AAdS spacetimes.\footnote{This
condition on the asymptotic Riemann was considered for first time
in \cite{ACOTZ4} in order to have a finite action principle in
even dimensions. It can be shown, however, that this asymptotic
behavior is also implied by the Fefferman-Graham expansion for
that tensor \cite{OleaJHEP}.} Then, the second line is
proportional to
\begin{equation}
\varepsilon _{abcd}\,\varepsilon ^{i_{1}i_{2}i_{3}i_{4}}\left[
\delta K_{i_{1}}^{j}e_{j}^{a}e_{i_{2}}^{b}+\delta
e_{j}^{a}e_{l}^{b}\left(
K_{i_{1}}^{j}\delta _{i_{2}}^{l}-K_{i_{2}}^{l}\delta _{i_{1}}^{j}\right) %
\right] \left(
R_{i_{3}i_{4}}^{cd}-K_{i_{3}}^{c}K_{i_{4}}^{d}+\frac{1}{\ell
^{2}}\,e_{i_{3}}^{c}e_{i_{4}}^{d}\right)\,,
\end{equation}
that is again cancelled by taking the boundary condition
(\ref{deltaK0},\ref{Kdelta}). This shows that, on the contrary to
standard conditions (\ref{Abarz},\ref{barAz}) for $3D$ AdS
gravity, the condition (\ref{Kdelta}) can indeed be lifted to
higher odd-dimensional AdS gravity. The boundary term derived from
this action principle equally cancels
the infinities in the Euclidean action and conserved quantities \cite%
{MOTZodd}.

The action principle for three-dimensional AdS gravity presented
here agrees with the Dirichlet problem up to a topological
invariant at the boundary. Even though this observation is almost
trivial in
three dimensions, we could expect that the boundary terms in ref.\cite%
{MOTZodd} (for $D=2n+1$) and ref.\cite{OleaJHEP} (for $D=2n$) generate the
full series of standard counterterms carrying out a suitable expansion.

%%%%%%%%%%%%%%%%%%%%%%%%%%%%%%%%%%%%%%%%%%%%%%%%%%%%%%%%%%%%%%%%%%%%%

\section*{Acknowledgments}

We wish to thank M. Ba\~{n}ados and S. Theisen for helpful
discussions. This work was partially funded by the grants 3030029
and 3040026 from FONDECYT (Chile) and by Funda\c{c}\~{a}o para a
Ci\^{e}ncia e Tecnologia (FCT) of the Ministry of Science,
Portugal, through project POCTI/ FNU/44648/2002. O.M. is also
supported by PUCV through the program Investigador Joven 2006.

%%%%%%%%%%%%%%%%%%%%%%%%%%%%%%%%%%%%%%%%%%%%%%%%%%%%%%%%%%%%%%%%%%%%%%%%%

\end{document}